\documentstyle[preprint,aps,tighten,epsf,floats]{revtex}
\newcommand{\ben}{\begin{displaymath}}
\newcommand{\een}{\end{displaymath}}
\newcommand{\be}{\begin{equation}}
\newcommand{\ee}{\end{equation}}
\newcommand{\bea}{\begin{eqnarray}}
\newcommand{\eea}{\end{eqnarray}}

\begin{document}
\draft
\title{Renormalization of QCD Coupling Constant in Terms of Physical Quantities}
\author{ G.Sh.Japaridze${ }^a$ and K.Sh.Turashvili${ }^b$ }
\address{${ }^a$ Department of Physics, Center for Theoretical studies of Physical
Systems, \\ Clark
Atlanta University, Atlanta, GA 30314
\\
${ }^b$ Institute for High Energy Physics, Tbilisi, Georgia}
\date{\today}
\maketitle

\begin{abstract}
A renormalization scheme is suggested where QCD input parameters - quark mass and coupling 
constant - are expressed in terms of gauge invariant and infrared stable quantities. For the 
renormalization of coupling constant the quark anomalous electromagnetic moment is used;
the latter is calculated in a two loop approximation. Examination of the 
renormalized $S$ matrix indicates confinement phenomenon already in the framework of 
perturbation theory. 
\end{abstract}

\pacs{
03.70.+k
12.39.Fe,
}
\newpage
Usually, in the charge and the mass renormalization procedure the so-called MOM scheme is used,
where the input parameters of the theory, appearing in action, are expressed in terms of 
on mass-shell Green's functions (Itzykson and Zuber, 1980; Ramond, 1989). Another widely accepted
scheme, MS, deals only with the divergent parts of the Green's functions, not appealing to any 
condition on momenta. Evidently, due to the renormalization invariance, any scheme is admissible.

In gauge theories like quantum electrodynamics (QED) and quantum chromodynamics (QCD) the input 
parameters are considered to be gauge invariant and infrared stable (i.e. not containing infrared 
divergences, generated by a massless gauge field). Let us for brevity call any quantity with these 
properties a physical quantity. Obviously, the observables, being measurable, should be the physical 
quantities but the converse may be not true (as an example, consider any function of field strength 
$F_{\mu\nu}$ in QED). Since Green's functions are not infrared stable and depend on a gauge (in 
general, their renormalization factors are nonlocal quantities (Basseto {\it et al}., 1987)), the 
interpretation of the results obtained from the schemes mentioned above may be obscured. Renormalized 
quantities may share undesirable properties, originated from the dynamics of the nonphysical 
degrees of freedom and the extraction of the physical information may be complicated.

Therefore, from our point of view, in gauge theories the most illuminating would be a scheme operating 
only with the physical degrees of freedom, thus allowing to avoid complications mentioned above. The 
prescription is as follows: calculate in regularized theory as many physical quantities as 
the input parameters are 
and express the latter in terms of physical quantities. So, in QED as well as in QCD, 
containig only two input parameters - coupling constant and the fermion mass (for 
illustrative purposes we consider here 
only one flavor) - we need the expressions for two physical quantities:
\begin{equation}
\mit\Sigma_{\alpha}=\mit\Sigma_{\alpha}(g_{0},m_{0};n);\mit\Sigma_{\beta}=
\mit\Sigma_{\beta}(g_{0},m_{0};n),
\end{equation}
where $g_{0}$ and $m_{0}$ are input parameters, the dependence of $\mit\Sigma$ on momenta, beinge 
irrelevant here, is omitted, $n$ is a space-time dimension and throughout this paper we use 
dimensional regularization. In QED and QCD $\mit\Sigma_{\alpha}$ and $\mit\Sigma_{\beta}$ have 
no limit at $n=4$ - regularization can not be removed in (1). Now resolve $g_{0}$ and $m_{0}$ in terms 
of $\mit\Sigma_{\alpha}$, $\mit\Sigma_{\beta}$ and $n$ and substitute into the expression for any 
other physical quantity $\mit\Sigma_{\gamma}$. In renormalizable theories, after all the calculations 
are performed, $\mit\Sigma_{\gamma}$ becomes finite in terms of $\mit\Sigma_{\alpha}$, 
$\mit\Sigma_{\beta}$ - regularization can be removed:
\begin{eqnarray}
\lim_{n\rightarrow 4}\mit\Sigma_{\gamma}(g_{0},m_{0};n)=
\lim_{n\rightarrow 4}\mit\Sigma_{\gamma}(g_{0}(\mit\Sigma_{\alpha},\mit\Sigma_{\beta};n),
m_{0}(\mit\Sigma_{\alpha},\mit\Sigma_{\beta};n);n)=\cr\cr
\lim_{n\rightarrow 4}\mit\Sigma^{\star}_{\gamma}(\mit\Sigma_{\alpha},\mit\Sigma_{\beta};n)
\equiv \sigma_{\gamma}<\infty
\end{eqnarray}
This scheme, describing the renormalization procedure as the expression of physical quantities in 
terms of physical quantities,\footnote{A similar procedure (in a different context) was 
discussed already in the early days of QED (Dyson, 1949).} allows to avoid any significant 
difficulties and seems most transparent from the physical point of view.

The aim of this paper is to develop such a scheme for QCD.

It is clear that we need two physical quantities. In (Tarrach, 1981) it was demonstrated that 
in the covariant gauge the solution of the equation
\begin{equation}
G^{-1}(p,m_{0},g_{0},\xi;n)\Psi_{in}(p)=0
\end{equation}
can be expressed in the framework of perurbation theory as
\begin{equation}
m=m_{0}+g_{0}^{2}\delta m_{1}(m_{0};n)+g_{0}^{4}\delta m_{2}(m_{0};n)+...,
\end{equation}
where $\delta m_{1}$ and $\delta m_{2}$ do not depend on a gauge parameter $\xi$ and are infrared 
stable. In (3), $G$ is a quark propagator and $\Psi_{in}(p)$ is the solution of the Dirac 
equation $(\gamma_{\mu}p_{\mu}-m)\Psi_{in}(p)=0$. The investigation in an axial gauge 
(Japaridze {\it et al}., 1991) confirms the gauge invariance and the infrared stability of $m$. 
In (Japaridze {\it et al}., 1991) it was shown that up to order $g_{0}^{6}$ the quark propagator 
has a simple pole at $p^{2}=m^{2}$. Therefore, the quark pole mass can be considered as on of the 
$\mit\Sigma$'s in relations (1) and to complete the scheme we have to point out the another physical
quantity in QCD.

We propose the quark anomalous electromagnetic moment, defined, as usual, from the amplitude of the 
quark elastic scattering on an external electromagnetic field:
\begin{equation}
\langle p|j_{\mu}|p+k \rangle A^{\mu}(k)=\bar{\Psi}_{in}(p)\Biggl( \gamma_{\mu}F_{1}(k^{2})+
\frac{i}{2}[\gamma_{\mu},\gamma_{\nu}]k^{\nu}F_{2}(k^{2}) \Biggr)\Psi_{in}(p+k)A^{\mu}(k)
\end{equation}
The anomalous electromagnetic moment $\chi$ is defined as $F_{2}(0)$.

We calculate $\chi$ up to order $g_{0}^{6}$ regularizing all the divergences (ultraviolet and 
infrared) in terms of space-time dimension $n$: 
\begin{equation}
\chi=g_{0}^{2}\chi_{1}(m_{0};n)+g_{0}^{4}\chi_{2}(m_{0};n),
\end{equation}
where $\chi_{i}$ is the $i$-loop contribution. The gauge dependent terms cancel in $\chi_{i}$; 
$\chi_{1}$ is infrared stable and the infrared divergence appears in $\chi_{2}$. To obtain the 
expression of order $g_{0}^{4}$, containig only $g_{0}$-associated divergences, we must use 
the relation (see (4))
$$
m_{0}=m-g_{0}^{2}\delta m_{1}(m_{0};n)+O(g_{0}^{4})=
m-g_{0}^{2}\delta m_{1}(m;n)+O(g_{0}^{4})
$$
in (6), i.e. reexpand the loop expressions:
\begin{eqnarray}
\chi=g_{0}^{2}\chi_{1}(m-g_{0}^{2}\delta m_{1}(m;n);n)+g_{0}^{4}\chi_{2}(m;n)=\cr\cr
g_{0}^{2}\chi_{1}(m;n)+g_{0}^{4}\Biggl( \chi_{2}(m;n)-
\delta m_{1}\frac{\partial \chi_{1}(m;n)}{\partial m} \Biggr)
\end{eqnarray} 
This reexpansion generates the infrared divergent tern $\partial \chi_{1}/\partial m$ which 
cancels the infrared divergent term in $\chi_{2}$.

So, $\chi$ is gauge invariante and infrared stable. Omitting the intermediary calculations, we quote 
the final result:
\begin{equation}
\chi=\frac{C_{F}g^{2}}{8\pi}\Biggl( 1-\frac{g^{2}}{8\pi^{2}}\biggl[ \frac{11C_{A}-2n_{f}}{3}
\Biggl( \frac{1}{n-4}+ln\;\frac{m^{2}}{4\pi\nu^{2}} \Biggr)+\Phi+O(g^{2}_{0};n-4) \biggr] \Biggr),
\end{equation}
where $C_{A}\equiv N$, $C_{F}\equiv (N^{2}-1)/2N$ are the invariants of $SU(N)$ group, 
$g^{2}\equiv g^{2}_{0}\nu^{n-4}$ is the dimensionless coupling constant, $\nu$ is the mass scale, 
appearing in the framework of dimensional regularization (Itzykson and Zuber, 1980; Ramond, 1989),
$n_{f}$ is the number of flavors, $m$ is the pole mass of the scattered quark, the external momenta 
obey $p^{2}=(p+k)^{2}=m^{2}$ and for the finite part $\Phi$ see the Appendix.

Thus from the point of view of renormalization procedure QED and QCD do not differ - in both of 
theories the input parameters can be expressed in terms of physical quantities. Surely, any scheme 
can be used but our goal was to demonstrate that it is possible to renormalize QCD coupling constant 
in terms of physical degrees of freedom. The scheme is described by the relations:
\begin{equation}
m_{0}=m_{0}(m,\chi;n),\;g_{0}=g_{0}(m,\chi;n)
\end{equation}
To use (9) we need the numerical values of $\chi$ and $m$. The gauge invariance and the infrared 
stability of these quantities does not mean necessarily that they are directly measurable, but 
these properties guarantee that the numerical values of $\chi$ and $m$ can be extracted from 
the experimental data. 
Of course, the numerical values of $\chi$ and $m$ depend not only on $m_{0}$ and $g_{0}$ but also 
on all other input parameters of, say, the Standard Model, but for the considered problem it is 
enough to analyze only QCD corrections.

The scheme (9) may be not suitable from the point of view of numerical convergence - the 
rate of convergence 
depends on the numerical values of $\chi$ and $m$. To improve the convergence, one has to 
introduce the so-called effective parameters (Itzykson and Zuber, 1980; Ramond, 1989) $g_{R}$ and 
$m_{R}$, i.e. to move to another scheme. It has to be pointed out that some statements formulated 
in terms of effective parameters (say, the increasing of the of the QCD effective coupling constant 
in the infrared region, sometimes interpretted as a physical effect of increasing the force between 
quarks at large distances) are scheme dependent and are not valid in another scheme. In other words, 
since the effective parameters are chosen arbitrarily, they can not affect the physical results.

To see this, let us illustrate how the renormalization group equation and the renormalization scheme 
arise in a quantum field theory. It is transparent in the framework of dimensional regularization, 
where we have two parameters $m_{0}$ and $g^{2}\equiv g^{2}_{0}\nu^{n-4}$. The mass scale $\nu$ 
defines the dimension of $g_{0}$, providing the dimensionless action. Parameters $g$ and $\nu$ are 
not independent:
\begin{equation}
g^{2}_{0}=g^{2}(\nu_{1})\nu_{1}^{4-n}=g^{2}(\nu_{2})\nu_{2}^{4-n},
\end{equation}
i.e.
\begin{equation}
\frac{dg^{2}(\nu)}{d\nu}+\frac{4-n}{\nu}g^{2}(\nu)=0
\end{equation}
The renormalization group equation (11) can be presented in a familiar form by introducing 
the effective parameter $g_{R}(\nu)$ by means of the relation
\begin{equation}
g^{2}=g^{2}(g^{2}_{R}(\nu),\nu)
\end{equation}
leading to
\begin{equation}
\nu\frac{dg^{2}_{R}(\nu)}{d\nu}=\beta(g^{2}_{R},\nu)
\end{equation}
where
\begin{equation}
\beta=\lim_{n\rightarrow 4}\frac{1}{{\partial g^{2}}/{\partial g^{2}_{R}}}\biggl[ (n-4)g^{2}-
\frac{{\partial g^{2}}}{{\partial \nu}} \biggr]
\end{equation}
The renormalization scheme is specified by the relation (12); then from (14) we obtain the appropriate 
$\beta$-function. For example, the choice
\begin{equation}
g^{2}=c_{1}(n)g^{2}_{R}(\nu)+c_{2}(n)g^{4}_{R}(\nu)+...
\end{equation}
results in
\begin{equation}
\beta=\lim_{n\rightarrow 4}\;(n-4)g^{2}_{R}\Biggl( 1-\frac{c_{2}(n)}{c_{1}(n)}g^{2}_{R}+... \Biggr)
\end{equation}
The schemes are labelled by particular choices of $c_{i}$ (e.g MS is obtained if we choose 
$c_{i}=a_{i}/(n-4)$). The introduction of $m_{R}(\nu)$ is based on the same argumentation.

So, the behavior of the effective charge defined through any particular scheme (e.g. leading to 
asymptotic freedom), being scheme dependent, may not lead to any valuable results - the behavior 
and numerical values of effective parameters depend on our choice and are not defined from the 
theory alone. 

Thus, the advantage of scheme (9), besides gauge invariant and infrared stability is that it does 
not operates with the effective parameters, allowing us to avoid conclusions which are insignificant 
from the physical point of view.

As it becomes evident, the behavior of $g_{R}$ and $m_{R}$ does not lead to a scheme-independent 
statements, e.g. the absence of quarks and gluons in the asymptotic states. On the other hand, 
from the renormalizability ofQCD it follows that in the expansion of any physical quantity 
$\sigma_{\gamma}$ (say, Wilson Loop)
\begin{equation}
\sigma_{\gamma}=\sum^{\infty}_{j=0}\chi^{j}\sigma_{\gamma,\;j}
\end{equation}
the coefficients $\sigma_{\gamma,\;j}$ are gauge invariant, infrared stable and contain no 
ultraviolet divergences at $n=4$, i.e. $\sigma_{\gamma,\;j}$ are finite.

So, the following question arises: are the quark and gluon degrees of 
freedom observable, i.e. do quarks and gluons appear in asymptotic states?

The answer may be obtained from the examination of the scattering matrix. We suggest the following 
criterion: if at least one $S$-matrix element built up in terms of fields is finite, the 
appropriate quanta appear in asymptotic states as a particles.

In QED it is well known that the electron elastic sacttering amplitude is infrared divergent 
and taking into account the emission of photons leads to the cancellation of these infrared 
singularities - only the inclusive cross sections are finite (Itzykson and Zuber, 1980; Ramond, 1989). 
That is why we can say that from the Dirac-Maxwell equations it follows the 
existence of electrons and photons as 
an observable particles. Of course, we {\it a priori} know that they exist and the $S$-matrix 
analysis is in accordance with the experimental data.

We consider the $S$-matrix element of scattering of quark on an external electromagnetic field. 
Note first that the existence of a physical quantity $\chi\equiv F_{2}(0)$ (see (5)) does not mean 
at all that the amplitude of the elastic sacttering is finite - in a full analogy with QED, 
the infrared divergences remain in the elastic amplitude after the charge and the mass renormalization 
is performed. Let us consider the gluon emission, assuming that the inclusive cross section might 
be finite. According to the LSZ reduction technique (Itzykson and Zuber, 1980; Ramond, 1989), for a 
gluon with the momentum $q$ and the polarization $\epsilon_{\mu}(q)$ in an asymptotic state we 
have
\begin{equation}
\frac{\epsilon_{\rho}(q)}{Z_{3}^{1/2}}q^{2}{\cal D}_{\mu\rho}(q)=
\frac{\epsilon_{\mu}(q)}{Z_{3}^{1/2}}q^{2}\frac{Z_{3}}{q^{2}}=Z_{3}^{1/2}\epsilon_{\mu}(q),
\end{equation}  
where $\cal{D}_{\mu\rho}$ is the gluon propagator and $Z_{3}^{1/2}$ is the gluon wave function 
renormalization factor. In order $g^{2}$ the contribution of the gauge field in the residue $Z_{3}$ 
is (Itzykson and Zuber, 1980; Ramond, 1989)
\begin{equation}
Z^{A}_{3}(q^{2})=iC_{A}\frac{g^{2}\nu^{4-n}}{2^{n+1}\pi^{n/2}}\;\frac{3n-2}{n-1}\;
\frac{\Gamma^{2}(n/2-1)\Gamma(2-n/2)}{\Gamma(n-2)}\;(q^{2})^{(n-4)/2},
\end{equation}
where $\Gamma$ is the Euler's function (Bateman and Erdelyi, 1973). The factor $Z_{3}$ contributes to 
the QCD coupling constant renormalization. To obtain the expression for the amplitude we have to 
perform all the calculations before the regularization is removed: renormalize $m_{0}$ and $g_{0}$ 
using (9), use (as we did for the amplitude (5)) the conditions $p^{2}=(p+k)^{2}=m^{2}$, $q^{2}=0$ 
and then set $n=4$. The condition $q^{2}=0$ means that $Z^{A}_{3}(0)$ can be equated to zero. This is 
analogous to the procedure of vanishing 
of integrals corresponding to a tadpole-type diagrams in a massless theory- because of the 
analyticity of theory in $n$ we can always find the region where the result is well-defined and 
then continue analytically in the desired region of $n$ (Itzykson and Zuber, 1980; Ramond, 1989). So
\begin{equation}
Z_{3}(q^{2})=Z^{A}_{3}(q^{2})+Z^{fermion}_{3}(q^{2})\rightarrow Z^{fermion}_{3}(0)
\end{equation}
at $q^{2}\rightarrow 0$. This means that only part  of the $g_{0}$ renormalization constant, namely 
$C_{A}/2-2n_{f}/3$ is restored (compare to $(11C_{A}-2n_{f})/3$ in expression (8)). In other words, 
$Z_{3}(0)$ 
does not renormalize $g_{0}$.

Therefore, if we consider the gluon emission in order to cancel the infrared divergences of the 
elastic scattering amplitude,  the ultraviolet divergence appears and the cross section 
in order $g^{4}$ contains an unavoidable infinity. The $S$-matrix is the same in any scheme but 
the result is transparent while using the scheme (9), manipulating only with the physical degrees of
freedom and not using the effective parameters.

Though the discussuin above is not a proof, it can be considered as an indication on a confinement 
phenomenon already in the framework of perturbation theory. In other words, QCD might be an 
example of a field theory where fields do not necessarily refer to the physical 
particles \footnote{ Long 
time ago Schwinger (Schwinger, 1962), starting from the idea that fields are more fundamental 
entities than particles, realized this conjencture in a two-dimensional electrodynamics, where the 
particles, corresponding to fermionic degrees of freedom, are absent in asymptotic states.}.

The next step would be the consideration of the scattering of colorless bound states. At the 
present time we have no definite result for this problem.

To conclude, in QCD it is possible to define renormalization procedure operating only in the space 
of physical degrees of freedom. Though the relations (9) guarantee that the physical quantities built 
up in terms of quark and gluon fields are finite, this is not enough for the existence of these field
quanta as physical particles. The examination of the quark scattering amplitude indicates that 
quark and gluon field quanta do not appear in asymptotic states.
\begin{center}
AKNOWLEDGMENTS
\end{center}
We would like to thank A. Bassetto, D. Finkelstein, J. Gegelia, C. Handy, N. Kiknadze, A. Khelashvili
and L. Vachnadze for the fruitful and illuminating discussions.
\newpage
\appendix
\section{}
The finite part $\Phi$ is
\begin{eqnarray}
\Phi=C_{F}\Biggl( -\frac{55\pi^{2}}{9}-8\gamma^{2}_{E}+2\pi^{2}ln\;2-3\zeta(3)+\frac{493}{12} \Biggr)
-\frac{2n_{f}\gamma_{E}}{3}+\frac{49n_{f}}{18}-22+\cr\cr
\frac{26\pi^{2}}{9}+4\gamma^{2}_{E}+
C_{A}\Biggl( \frac{131\pi^{2}}{18}+\frac{1675\gamma^{2}_{E}}{108}+
\frac{17\gamma_{E}}{9}-\pi^{2}ln\;2+\frac{3}{2}\zeta(3)-\frac{959}{72} \Biggr)-
\frac{\pi^{2}}{2}\bigtriangleup^{1/2}_{i}-\cr\cr
\bigtriangleup_{i}(6+4ln\;\bigtriangleup_{i})+
\frac{5\pi^{2}}{2}\bigtriangleup^{3/2}_{i}+\cr\cr
\bigtriangleup^{2}_{i}\biggl[ 4\gamma^{2}_{E}-
\frac{2\pi^{2}}{3}-3ln^{2}\;\bigtriangleup_{i}-8+(\frac{4}{9}ln\;\bigtriangleup_{i}-\frac{20}{27})F_{i1}
 +\frac{4}{9}F^{\prime}_{i1}+(\frac{4}{9}ln\;\bigtriangleup_{i}-\frac{38}{27})F_{i2}+
\frac{4}{9}F^{\prime}_{i2} \biggr]\cr\cr+
\bigtriangleup^{3}_{i}\biggl[ (\frac{2}{3}ln\;\bigtriangleup_{i}-\frac{11}{9})F_{i3}+
\frac{1}{3}F^{\prime}_{i3}+(3-2ln\;\bigtriangleup_{i})F_{i4}-2F^{\prime}_{i4} \biggr],
\end{eqnarray}
where $m$ is the mass of the scattered quark, $\bigtriangleup_{i}\equiv m^{2}_{i}/m^{2}$, 
$m_{i}\neq m$,\\ $\gamma_{E}\simeq 0.5771$ is the Euler constant and
\be
\zeta(3)=\sum^{\infty}_{j=1}\frac{1}{j^{3}}
\label{zeta}
\ee
is the Riemann zeta function.

The $F_{ij}$'s are the generalized hypergeometric functions (Bateman and Erdelyi, 1973) 
of $\bigtriangleup_{i}$ and the parameters of $F_{ij}$ are as follows:
\begin{equation}
F_{i1}=_{3}F_{4}\pmatrix{ 1,\frac{n+2}{2},\frac{n-3}{2},\frac{n-2}{2}\cr 
2,n-2,\frac{n+1}{2}\cr},
\end{equation}
\begin{equation}
F_{i2}=_{2}F_{3}\pmatrix{ 1,\frac{n-1}{2},\frac{n-2}{2}\cr
\frac{n+1}{2}, \frac{n}{2}\cr},
\end{equation}
\begin{equation}
F_{i3}=_{3}F_{4}\pmatrix{ 1, \frac{n+2}{2},\frac{n-1}{2},\frac{n-2}{2}\cr
\frac{n+1}{2}, 3, n-1\cr},
\end{equation}
\begin{equation}
F_{i4}=_{1}F_{2}\pmatrix{ 1, \frac{n-2}{2}\cr
\frac{n+2}{2}\cr}
\end{equation}
and 
\begin{equation}
F^{\prime}_{ij}\equiv \frac{dF_{ij}}{dn}
\end{equation}
The functions $F_{ij}$ and $F^{\prime}_{ij}$ are considered at $n=4$. 
\newpage
\begin{center}
REFERENCES
\end{center}
Bateman, H. and Erdelyi, A. (1973). {\it Higher Transcendental Functions, McGraw-Hill, New York}.\\
Bassetto, A., Dalbosco, M. and Soldati, R. (1987).{\it Physical Review D}, {\bf 36}, 3138.\\
Dyson, F.J. (1949).{\it Physical Review}, {\bf 75}, 1736.\\
Itzykson, C. and Zuber, J.B. (1980). {\it Quantum Field Theory, McGraw-Hill, New York}.\\
Japaridze, G., Khelashvili, A. and Turashvili, K. (1991). {\it In Proceedings of XII International 
Conference "Quarks-90", World Scientific, Singapore}, 147.\\
Ramond, P. (1989). {\it Field Theory: A Modern Primer, Addison-Wesley, Reading, Massachusetts}.\\
Schwinger, J. (1962). {\it Physical Review}, {\bf 128}, 2425.\\
Tarrach, R. (1981). {\it Nuclear Physics B}, {\bf 183}, 384.\\
\end{document}